\documentstyle[amssymb,aps,twocolumn]{revtex}

\begin{document}
\title{Low-Temperature Glassy Response of Ultrathin Manganite Films to Electric and
Magnetic Fields}
\author{A. Bhattacharya, M. Eblen-Zayas, N. E. Staley, A. L. Kobrinskii, and A. M.
Goldman}
\address{School of Physics and Astronomy, University of Minnesota, 116 Church St. SE,%
\\
Minneapolis, MN 55455, USA}
\date{
\today%
}
\maketitle

\begin{abstract}
The glassy response of thin films of La$_{0.8}$Ca$_{0.2}$MnO$_{3}$ to
external magnetic and gated electrostatic fields in a field-effect geometry
has been studied at low temperatures. A hierarchical response with
irreversible memory effects, non-ergodic time evolution, aging and annealing
behavior of the resistance suggest that the dynamics are governed by strain
relaxation for both electronic and magnetic perturbations. Cross-coupling of
charge, spin and strain have been exploited to tune the coercivity of an
ultrathin manganite film by electrostatic gating.
\end{abstract}

\pacs{PACS numbers: ()}

Manganites, known for their `colossal' magnetoresistance (CMR), possess a
diversity of phases driven by correlations between the spin, charge and
orbital degrees of freedom of the electrons and their strong coupling to the
lattice \cite{Millis}. Localized electrons on Mn$^{3+}$ sites create large
lattice distortions via the Jahn-Teller effect, causing strong strain fields
to develop \cite{Renner}. When these electrons are delocalized, for example
by the double exchange mechanism between aligned Mn core spins, the local
strain is relieved. Under appropriate circumstances, manganites have an
admixture of phases of very different electronic, magnetic and structural
properties, but nearly equal free energies \cite{Burgy}. Consequently, the
properties of these systems may be strongly susceptible to external
perturbations that lead to phase conversion within the admixture \cite
{Millis1} , giving rise to `colossal' effects. The presence of competing
strain fields, Coulomb interactions, magnetic correlations and defects may
frustrate this process, giving rise to a complex free energy landscape with
many nearly degenerate minima and hierarchical barriers. This naturally
gives rise to glass like dynamics \cite{Palmer,Uehara}, at low temperatures.
Our understanding of the response of such a system to external forces is
complicated by the cross-couplings between the different degrees of freedom.
However, if the hierarchy of barriers being crossed can be attributed to
just one degree of freedom (e.g. spin or strain), then that `rate limiting'
property governs the dynamics on large time scales, simplifying the
analysis. Conversely, cross-couplings present the opportunity to influence
one kind of order with a force that couples to a different variable.

In this letter we investigate the response of ultrathin films of La$_{0.8}$Ca%
$_{0.2}$MnO$_{3}$ (LCMO) to electric and magnetic fields. This composition
is close to the phase boundary between a ferromagnetic metal (FM) at higher
Ca doping and a ferromagnetic charge ordered insulator (F-COI) at lower
doping. Bulk single crystals of similar composition are believed to exist in
a mixed phase with coexisting regions of insulating and metallic properties 
\cite{Algarabel}, with the transport properties at low temperatures arising
from percolation of metallic regions. The samples are typically 21 u.c ($%
\backsim $82\AA ) thick films of La$_{0.8}$Ca$_{0.2}$MnO$_{3}$ grown using
ozone-assisted molecular beam epitaxy on surface treated SrTiO$_{3}$(STO)
substrates locally thinned to 35-50$\mu $m \cite{Bhattacharya}, permitting a
field-effect geometry. A Pt electrode (1000\AA\ thick) on the back of this
thinned region serves as the gate. The manganite film is patterned into a
wire 100$\mu $m wide, with tabs for carrying out four terminal measurements,
which were performed using standard DC techniques. The gate-drain current
was always monitored, and remained below 0.6nA, while the source-drain
measurement current was 100nA.

We observe a magnetic transition at about 150K, and an accompanying
resistive transition, from an activated insulating state to a nominally
metallic state near the Curie temperature, along with CMR [Fig.1]. However,
at the lowest temperatures (%
\mbox{$<$}%
36K) there is a reentrant insulating phase \cite{Ziese}. Near the resulting
minimum, the resistance has a large susceptibility to gate and magnetic
fields. We observe clear signatures of hierarchical energy barriers, glassy
dynamics and aging in the response. We argue that the dynamics are governed
by structural relaxation. Cross-coupling has been exploited to effect a
measurable change in the magnetic coercivity of a sample upon application of
a gate electric field.

The details of the gate effect at low temperatures will be discussed
elsewhere \cite{Eblen-Zayas}. For the present purpose, it suffices to note
that the gate electric field couples to the charge degrees of freedom, while
the magnetic field couples to spin. The applied gate voltage is always
negative, inducing `hole' like charge carriers. The electrostriction in STO
at low temperatures is known to saturate at electric fields of approximately
15kV/cm \cite{Bhattacharya}, well below 85kV/cm, the maximum field applied
here. Thus, the effects that we observe are not due to biaxial substrate
strain.

Within a mixed phase scenario, measurements of resistance are particularly
sensitive to changes in the percolative metallic path. Upon applying an
external field that favors the growth of one phase with respect to the
other, the change in resistance depends on the motion of domain boundaries
separating the two. In this context, the presence of hierarchical energy
barriers was graphically demonstrated \cite{Levy} in resistivity
measurements of La$_{0.5}$Ca$_{0.5}$Mn$_{0.95}$Fe$_{0.05}$O$_{3}$. We have
measured a similar hierarchical response to both applied electric and
magnetic fields at 30K. A succession of external magnetic/gate electric
field pulses of different values and duration were turned on and off for
times of 0.5-1.5hr and 1hr, respectively. When a field was turned on, the
resistance decreased to a value $R_{ON}$, with a large `fast' change and a
smaller `slow' part of that evolved with a logarithmic dependence on time.
On turning off the field, the resistance relaxed to an intermediate value $%
R_{TR}$ (thermo-remnant resistance), indicating that the sample resistance
undergoes an irreversible change, analogous to thermo-remnant magnetization
in spin glasses \cite{Vincent}. On subsequent pulses, if the value of the
field exceeded that of the maximum field previously applied, $R_{TR}$
decreased further, but otherwise the response was reversible and $R_{TR}$
did not change [Fig.2(a,b)]. Thus, the $R_{TR}$ has memory of the previous
highest field applied. Furthermore, the $R_{TR}$ does not discriminate
between the application of an electric or magnetic field. This was
demonstrated by first turning on a large gate electric field, turning it
off, and then applying a series of increasing magnetic field pulses
[Fig.2(c)]. The hierarchy of barriers for the lowering of resistance due to
an applied magnetic field seems to `respect' the barriers already crossed by
application of the gate electric field, regardless that they couple to spin
and charge respectively.

We interpret the hierarchical barriers in terms of the dynamics of pinned
domain walls that separate the insulating and metallic regimes. The mutual
equivalence of barriers can be explained if they are primarily due to strain
in the insulating patches. Competing strain fields can cause frustration and
pinning of the domain walls \cite{Ahn}. On application of an external force,
pinning sites up to a certain threshold are overcome and the walls move
irreversibly. The strain field acquires a new configuration and can now
relax in a reversible manner in response to external forces until a pinning
site of the next higher strength is encountered. A greater external force is
required to bring about the next irreversible change. If the hierarchy of
pinning is determined by the strain fields alone, it does not matter that
the force is applied via aligning spins or inducing charge at the domain
walls since they cross-couple to the same strain field.

We also observe thermally assisted crossing of hierarchical barriers, or
`annealing'. Upon warming the film from the lowest temperature up a
temperature T$_{anneal}$, with the gate/magnetic field on, and cooling back
down, the resistance of the film changes irreversibly. On subsequent thermal
cycles, the resistance varies reversibly so long as T$_{anneal}$ of the last
cycle is not exceeded. The qualitative similarity between the annealing
curves in electrostatic and magnetic fields again suggests a common
mechanism. This also bears a striking resemblance to the evolution of the
resistance seen upon structural annealing of amorphous quench-condensed
films [Fig. 3(c)] \cite{Silverman}. This is due to crystallization at ever
increasing length scales as the film is progressively annealed at higher
temperatures, resulting in a `mixed phase of amorphous and crystalline
material .

We now turn to the dynamics of the response. On turning on a gate electric
field or magnetic field, a small part of the response evolves
logarithmically in time [Fig.4(a)], typical of glasses. We also observe wait
time dependence or aging of the relaxation in a magnetic field. After
cooling the sample in zero field to 30K, a large magnetic field of 2.5T was
turned on at 30K for about 1hr. The field was then set to zero and the
sample allowed to relax for 1hr to define an appropriate base line value for
the resistance. Subsequently, a smaller field of 1.5T was turned on for
different wait times $t_{W}$ [Fig.4(c)] and then turned off, and the sample
was allowed to relax for approximately $3t_{W}$. The slow logarithmic
relaxation that occurs after this field is turned off depends inversely upon 
$t_{W}$. The curves for wait times of up to 120min collapse when the
relaxation times are scaled by $t_{W}$ [Fig.4(d)]. This is consistent with
ideas about aging in the context of spin and structural glasses \cite
{Vincent}, where $t_{W}$ determines the height of the largest energy barrier
crossed by the system, and consequently the rate limiting relaxation time
scale when the field is turned off. The rate of relaxation was observed to
slow down to the point of saturation between wait times of 120min and
240min, shown by the fact that these relaxation curves fall on top of one
another within the spread of our data. This may indicate that the phase
space for barriers of incrementally higher value is very limited for the
1.5T magnetic field applied. We have also measured wait time dependence with
a gate electric field. With the application of the highest fields available (%
$300V=85kV/cm$), aging was not observed to within the uncertainty of our
data, and the relaxation traces seem to lie on top of each other[Fig.4(b)].
This is not merely due to the size of the gate electric field, since we do
observe aging in a magnetic field with the same wait times and equivalent
strength (1T). This is also in contrast to the aging observed on gating
charge glasses \cite{Vaknin} and may be due to the fundamentally different
ways that gate and magnetic fields affect the film. In the insulating
regions, the gate electric field is not screened, and goes right through the
film. In metallic regions, it is strongly screened and only affects the
first few unit cells at the dielectric-film interface. The strongest effects
are likely to be felt near the insulator/metal boundaries, where the
electric field would couple maximally to the film \cite{Wu}. In contrast, an
applied magnetic field would influence all the spins in the film
equivalently. The relaxation of strain occurs throughout the insulating
regions, as opposed to just at the boundaries for the gate effect. Thus,
during $t_{W}$, the system is able to sample a greater phase space in the
hierarchy of higher energy barriers during the application of a magnetic
field, producing a more pronounced slowing down or aging effect. However,
since the barriers being crossed just at the boundaries are the same in
magnitude, the hierarchy is preserved in resistance measurements.

As a further example of strong cross-couplings, we have measured the effect
of a gate electric field on the magnetic moment of a film, motivated by
recent work on magnetic semiconductors \cite{Ohno}. In manganite films below
a certain thickness, $H_{C}$ is enhanced both by decreasing thickness and
lowering temperature \cite{Steren}, similar to the behavior found in AuFe
cluster glasses \cite{Campbell}. This is attributed to the pinning of
magnetic domain walls in a metastable energy landscape, due to defects
caused by strain or other microstructure \cite{Gaunt}.

We used an unpatterned film (21u.c.) with a large area gate (12 mm$^{2}$) to
allow detection of small changes in the magnetic moment of the film in a
SQUID magnetometer. The coercivity at 2K was 484Oe, significantly higher
than that of thicker films of the same composition. The sample was cooled in
zero field and then saturated by applying +5000 Oe. The field was then
reversed to --450 Oe, a gate voltage of --200V (-30kV/cm) was applied across
the dielectric in 100V steps, and the moment was measured as a function of
time [Fig.5]. The sign of the total moment reversed, indicating that $H_{C}$
was crossed. Considering that the effective gated area was only about 55\%
of the total film area, the change in magnetic moment of the gated area was
about 36\% of the saturation value. Furthermore, upon cooling the film in a
5.5T field from above Tc to 2K, $H_{C}$ was reduced by about 120Oe compared
to the ZFC value. Since application of either a gate or magnetic field
serves to lower the resistance, our measurements indicate that the pinning
of magnetic domain walls is correlated to the fraction of insulating phase
present in our film. It is likely that the domain walls are pinned by
strained regions of insulating material, much as inclusions and strain do in
other systems. The reduction in H$_{C}$ is then related to the lowering of
the number and strength of pinning sites upon reducing the fraction of the
insulating phase.

This work was supported by the National Science Foundation under grant
NSF/DMR-0138209 and by the University of Minnesota MRSEC (NSF/DMR- 0212032)$%
. $ M. Eblen-Zayas was supported by an NSF Graduate Research Fellowship.


\begin{figure}[tbp]
\caption[Evolution of$R(T)$ of the 9.19\AA\ film as a function of in-plane
magnetic field. Field values from top to bottom are: 12.5, 12, 11.6, 11.5,
11, 10, 9, 8, 7, 6, 5, 4, 3, 2, and 0 T. Inset: temperature at which dR/dT
becomes zero is plotted vs. applied field.]{Hierarchical response: (a) Upper
panel shows gate voltage ($V_{G}$) sequence of --100, -200, -50, -100, -300V
with intervening zeros, and lower panel shows the corresponding changes in
resistance (b) Sequence of magnetic fields, of 0.2, 0.4, 2, 0.4, 1.0, 3.0T
with intervening zeros and response (c) Equivalence of hierarchy: gate pulse 
$V_{G}$=-300V pulse followed by magnetic field pulses of 0.2, 0.4, 0.6 and
2.0T. }
\label{fig2}
\end{figure}
\begin{figure}[tbp]
\caption{ (a) Progressive annealing of LCMO film, with temperature, in a
gate electric field. The double sided arrows indicate reversible paths for
thermal cycling, while the single sided arrows indicate irreversible changes
and (b) similar effect in a magnetic field. (c) Annealing of an a-Bi film on
warming up (adapted from Ref\protect\cite{Silverman})}
\label{fig3}
\end{figure}
\begin{figure}[tbp]
\caption{Glassy dynamics and aging at 30K: (a) The response $R(t)$ after
turning on a magnetic field of 2.5T (200 Oe/s) and gate voltage of --300V,
after cooling in zero field to 30K. In the aging experiments, initially a
field of 2.5T was turned on for 1hr, then off for 1hr to make the subsequent
response reversible reversible at lower fields. (b) G(t) with gate field
pulses of -300V turned on for different $t_{W}$, and measured upon turning
off, showing no discernable signs of aging (c) $G(t)$ with magnetic field
pulses of 1.5T of different $t_{W}$. (d) Wait time scaling of $G(t/t_{W})$
from (c).}
\label{fig4}
\end{figure}
\begin{figure}[tbp]
\caption{Gate induced change in magnetic moment. A magnetic field of --450
Oe was applied at t=0 after saturating at 5000 Oe. The gate voltage was
changed in two steps. Inset (i) indicates section of hysteresis loop where
the gate was turned on, and (ii) is a zoomed in view of the change in
magnetic moment. The effect of the gate is equivalent to an abrupt increase
in magnetic field of 50 Oe.}
\label{fig5}
\end{figure}

\end{document}